\documentclass[letter,twocolumn]{jpsj3}
\newcommand{\uu}{\mib{u}}
\newcommand{\ee}{\mib{e}}
\newcommand{\kk}{\mib{k}}
\newcommand{\KK}{\mib{K}}
\newcommand{\qq}{\mib{q}}
\newcommand{\GG}{\mib{G}}
\newcommand{\bfl}{\mib{l}}
\newcommand{\bfm}{\mib{m}}
\newcommand{\Rl}{\mib{R_l}}
\newcommand{\Rm}{\mib{R_m}}
\newcommand{\Rlm}{\mib{R_{lm}}}
\newcommand{\Rd}{R_{\rm d}}
\newcommand{\ul}{\mib{u_l}}
\newcommand{\um}{\mib{u_m}}
\newcommand{\ulm}{\mib{u_{lm}}}
\newcommand{\uud}{\uu_{\rm d}}
\newcommand{\nee}{n_{\rm e}}
\newcommand{\kF}{k_{\rm F}}
\newcommand{\kB}{k_{\rm B}}
\newcommand{\Vion}{V_{\rm ion}}
\newcommand{\Vel}{V_{\rm el}}
\newcommand{\qt}{{\qq{\rm t}}}
\newcommand{\ql}{{\qq{\rm l}}}
\newcommand{\ct}{c_{\rm t}}
\newcommand{\cl}{c_{\rm l}}
\newcommand{\eql}{\ee_{\qq{\rm l}}}
\newcommand{\eqt}{\ee_{\qq{\rm t}}}
\newcommand{\eqlam}{\ee_{\qq \lambda}}
\newcommand{\ctzero}{c_{\rm t0}}
\title{Melting Temperature of Metals Based on the Nearly Free Electron Model}

\author{Tamifusa {\sc Matsuura}\thanks{Present address: Department of Physics, Nagoya  University, Chikusa-ku, Nagoya 464-8602, Japan.}, 
Hidenori {\sc Suzuki}\thanks{Present address: Department of Physics, College of Humanities and Sciences, Nihon University, 
Setagaya-ku, Tokyo 156-8550, Japan.}, 
Ken'ichi {\sc Takano}, and Fumihiro {\sc Honda}}
\inst{Toyota Technological Institute, 
Tenpaku-ku, Nagoya 468-8511, Japan}

\abst{
We propose a general formula for the melting temperature of metals in terms of electronic mass, electronic number, and nearest-neighbor lattice distance. 
We derive it from the instability of the transverse phonon in the solid phase, using the nearly free electron model. 
Including higher order terms of vibrations enhanced near the melting temperature, the electronic restoring force is reduced and the ionic one is negligible. 
This fact greatly brings down the melting temperature, bringing it close to experimental data in the range of 10~\% for Cs, Cu, Au, and Ba. 
Also, this theory confirms the Lindemann criterion. 
}

\kword{melting temperature, melting of metals, nearly free electron model, Lindemann criterion}

\begin{document}
\maketitle

The continuing demand for composite metals to meet wider range 
characteristics requires a fundamental and more practical method. 
Among various physical quantities, melting temperature is one of the most fundamental ones. 
The principal requirement of a theory of melting is to calculate the Gibbs free energies $G_{\rm s}$ and $G_{\rm l}$ of the solid and liquid phases as functions of pressure $P$ and temperature $T$. 
The melting curve in the $P$-$T$ plane is then determined by the condition $G_{\rm s}(P,T)=G_{\rm l}(P,T)$, where the calculations of $G_{\rm s}(P,T)$ and  $G_{\rm l}(P,T)$ are taken as two separate problems. 
Statistical aspects of melting temperature have received the greatest share of attention. 
Modern computing techniques have made it possible to compare various approximate schemes with one another and with actual and computer experiments~\cite{Metropolis,Alder1}. 
Most studies have been based on inverse power law~\cite{Brush,Hansen1,Hoover2}, hard-core~\cite{Alder2,Wood,Alder3}, or other relatively idealized and short-range forces~\cite{Singh,Shapiro}, such as the Lennard-Jones force field~\cite{Hansen2}. 
We cite some of various attempts~\cite{Lou,Mulargia,Ida,Wilsdorf}. 

Stroud and Ashcroft studied melting phenomena in Na based on the electron gas model with the electron-ion and electron-electron interactions~\cite{Stroud}. 
They calculated the free energies $G_{\rm s}$ and $G_{\rm l}$ of the solid and liquid phases separately and determined the melting temperature by the condition of $G_{\rm s}=G_{\rm l}$. 
After elaborate calculations, they obtained the melting curve which was claimed to be in good agreement with the results of experiments, up to at least 40 kbar in Na. 
The calculation however is not suitable for other metals except alkali metals, since in the calculation the properties of the free electron were  fully taken into account. 

Another approach is to determine the melting temperature by the instability of the solid phase. 
Lindemann investigated the instability condition of the solid phase and proposed a criterion for the melting temperature that the Lindemann ratio $\delta\sim 0.1$ where $\delta $ is the ratio of the mean square amplitude of vibration of each atom about its lattice site to the nearest-neighbor distance of the lattice sites~\cite{Lindemann}. 
Using the Lindemann criterion, we can obtain the melting temperature of any crystal. 
Born calculated the melting temperature from the vanishing point of  an elastic stiffness constant $c_{44}$, which means that the instability of the shear vibration in the solid phase occurs at the melting temperature~\cite{Born}. 
He confirmed the Lindemann criterion in the solid phase with the Lennard-Jones potential. 
Fukuyama and Platzman calculated the transverse mode instability point using the self-consistent harmonic approximation (SCHA) as the onset of the superheating transition. They also applied the SCHA to derive Lindemann criterion for the alkali metals and to estimate the melting density and temperature of a Coulomb solid.~\cite{Fukuyama, Platzman}
These theories, however, may not be suitable for metals.
Metals are the most plastic solid, 
and the cohesive energy is mainly a function of density of packing. Local deviations from a strict lattice regularity are easily accommodated~\cite{Ziman}. 
Actually, conduction electrons derive an electronic potential of  long-range and oscillating character through the adiabatic principle. 

In this paper, we construct a general formula for the melting temperature as the vanishing point of $\ct$, which is the velocity  of the transverse phonon in metals. 
To calculate $\ct$, we use the SCHA for lattice vibrations and 
the nearly free electron model for conduction electrons. 
Finally, the melting temperature is given as 
%(1)---------------------------------------------
\begin{align}
T_{\rm m} = 0.145009 \, \times \, 
\frac{\hbar^2 \nee}{m^*\Rd^2\kB} 
\label{melting_temp}
\end{align} 
%---------------------------------------------
with the Boltzmann constant $\kB$, the Plank constant $\hbar$, and the nearest-neighbor lattice distance $\Rd$. 
The parameters $m^*$ and {$\nee$} are the effective mass and the number of conduction electrons per site. 
Except for the numeric factor 0.145009, 
this equation is on the order of the Fermi temperature 
$\hbar^2/(m^* \Rd^2\kB)$$\sim$$\hbar^2\kF^2/(2m^* \kB)$ with the Fermi momentum $\hbar \kF$. 
The numeric factor is brought by the procedure beyond the harmonic approximation. 
Note that eq.~(\ref{melting_temp}) does not include the ionic mass $M$. 

The melting temperatures calculated using eq.~(\ref{melting_temp}) agree well with the experimental ones for alkali and noble metals. 
Moreover, the melting temperatures of various pure and composite metals can be estimated using the parameters as easily accessible experimental data in the literature. 
This theory also deduces the Lindemann ratios of 0.183 and 0.172 for the bcc and fcc lattices, respectively. 
These values are consistent with the Lindemann criterion. 

We proceed to microscopically derive the formula (\ref{melting_temp}) for the melting temperature of metals.  
Atoms in a metallic material vibrate about their equilibrium positions. 
We denote the equilibrium position of the atom at the site $\bfl$ as $\Rl$ and the deviation as $\ul$. 
Then, the ionic potential $\Vion$ is written as 
%(2)---------------------------------------------
\begin{align}
\Vion &= \frac{1}{2}\sum_{\bfl,\bfm}v(\Rlm+\ulm) 
\label{ionic_pot}
\end{align}
%---------------------------------------------
with $\Rlm \equiv \Rl-\Rm$ and $\ulm \equiv \ul-\um$, where $v(\Rlm+\ulm)$ is the ionic potential between the atoms at the sites $\bfl$ and $\bfm$. 
In the adiabatic approximation, the one-electron Hamiltonian of the conduction electrons is 
%(3)---------------------------------------------
\begin{align}
H_{\rm el} 
&= -\frac{\hbar^2}{2m} \nabla^2 + \sum_{\bfl}w({\bf r}-\Rl-\ul) , 
\label{one_p_Hamiltonian}
\end{align}
%---------------------------------------------
where $m$ and $w({\bf r}-\Rl-\ul)$ stand for the electron mass and the potential of the atom at the site ${\bfl}$ with the coordinate $\Rl+\ul$, respectively. 
The electron field is represented as  
$\psi_{\sigma}({\bf r})=\sum_{\bfl} a_{\bfl\sigma} \phi({\bf r}-\Rl-\ul)$, where $\phi({\bf r}-\Rl-\ul)$ is the Wannier function at the site $\bfl$ and $a_{\bfl\sigma}$ is the annihilation operator of the conduction electron at the site $\bfl$ with the spin $\sigma$. 
The electronic potential $\Vel$ 
is written as 
%(4)---------------------------------------------
\begin{align}
\Vel 
&=\left\langle \sum_{\sigma} \int \psi_{\sigma}^*({\bf r}) 
H_{\rm el} 
\psi_{\sigma}({\bf r}) d{\bf r}\right\rangle_{\rm el} \nonumber\\
&= \frac{1}{2}\sum_{\bfl,\bfm,\sigma}t(\Rlm+\ulm) 
\left( \langle a_{\bfl\sigma}^\dagger a_{\bfm\sigma}\rangle _{\rm el}+{\rm h.c. } \right) , 
\label{electronic_pot}
\end{align}
%---------------------------------------------
where $t(\Rlm+\ulm)$ $\equiv$ $\int \phi^*({\bf r}-\Rl-\ul)$$H_{\rm el} 
$$\phi({\bf r}-\Rm-\um)d{\bf r}$ 
is the transfer integral between conduction electrons at the sites $\bfl$ and $\bfm$. 
$\langle\cdots \rangle_{\rm el}$ denotes the thermal average over electronic distributions in the ionic configuration $\{\Rl+\ul\}$.
Thus, the Hamiltonian of lattice vibrations is 
%(5)---------------------------------------------
\begin{align}
H = \frac{M}{2}\sum_{\bfl} \left( \frac{d\ul}{dt} \right)^2 
+ \Vion + \Vel . 
\label{Hamiltonian}
\end{align}
%---------------------------------------------
In what follows, $\Vion^{(0)}$ and $\Vel^{(0)}$ respectively denote $\Vion$ and $\Vel$ with $\ul = 0$ for all $\bfl$ sites, 
and also their differences are $\Delta\Vion \equiv \Vion - \Vion^{(0)}$ and $\Delta \Vel \equiv \Vel - \Vel^{(0)}$. 
We note that $\{\Rl\}$ forms the minimum configuration of $\Vion^{(0)}$+$\Vel^{(0)}$; $\partial (\Vion^{(0)}+\Vel^{(0)})/\partial \Rl=0$. 

We show how to incorporate nonlinear terms to the restoring force and see how the procedure actually reduces the restoring force. 
First we calculate $\Delta\Vion$, which is expanded as 
%(6)---------------------------------------------
\begin{align}
\Delta \Vion 
=\frac{1}{2}\sum_{\bfl,\bfm}\sum_n\sum_{\alpha_1,\cdots,\alpha_n}
&\frac{1}{n!}\frac{\partial^n v}{\partial R_{\bfl\alpha_1}\cdots\partial R_{\bfl\alpha_n}}
\nonumber\\
& \times u_{\bfl\bfm\alpha_1}\cdots u_{\bfl\bfm\alpha_n},  
\end{align}
%---------------------------------------------
where ${\partial^n v}/{\partial R_{\bfl\alpha_1}\cdots\partial R_{\bfl\alpha_n}}$ is a derivative with  $\ulm = 0$ for all $\bfl$ and $\bfm$ sites, and the suffices $\alpha_1,\cdots, \alpha_n$ denote components. 
In the SCHA, the product $u_{\bfl\bfm\alpha_1}\cdots u_{\bfl\bfm\alpha_n}$ is  decoupled in pairs.  
For example, the terms of the same type as $u_{\bfl\bfm\alpha_1}^2 u_{\bfl\bfm\alpha_2}^2 u_{\bfl\bfm\alpha_3}^2$ reduce to 
${6 \choose 2}{4 \choose 2}
u_{\bfl\bfm\alpha_1}^2\langle u_{\bfl\bfm\alpha_2}^2\rangle \langle u_{\bfl\bfm\alpha_3}^2\rangle$, 
where $\langle \cdots\rangle $ denotes the thermal average over ionic configurations in the equilibrium and 
the prefactor is the number of combinations for decoupling. 
Then, we apply the Fourier transformation 
$v(\Rl)=(1/N_L)\sum_{\qq}v(\qq)e^{i\qq\cdot\Rl}$ 
with $N_L$, which is the number of lattice sites. 
Thus, $\Delta \Vion$ is written as 
%(7)---------------------------------------------
\begin{align}
\Delta \Vion =
&\frac{1}{4 N_L}\sum_{\qq} v(\qq)\sum_{\bfl,\bfm}(i\qq\cdot\ulm)^2 \nonumber\\
&\times e^{i\qq \cdot (\Rl-\Rm)}\exp\left[ -\frac{1}{2}\langle (\qq\cdot\ulm)^2\rangle \right] .
\end{align}
%---------------------------------------------
The nonlinear terms have been incorporated into the exponential factor $\exp[-\langle (\qq\cdot\ulm)^2\rangle/2 ]$ 
that reduces $\Delta \Vion$. 
Here, we introduce a common reduction factor  $\exp[-\langle (\qq\cdot\uud)^2\rangle/2 ] $, where  $\uud$ is the displacement between the nearest-neighbor sites, since the nearest-neighbor terms are dominant in the equation. 
Defining the effective interaction  $\bar{v} (\qq)=v(\qq)\exp[-\langle (\qq\cdot\uud)^2\rangle/2 ] $, we obtain 
%(8)---------------------------------------------
\begin{align}
\Delta \Vion = 
\frac{1}{4 N_L}\sum_{\qq} \bar{v} (\qq)\sum_{\bfl,\bfm}(i\qq\cdot\ulm)^2e^{i\qq \cdot (\Rl-\Rm)}. 
\label{del_v_c_exp}
\end{align}
%---------------------------------------------
We apply the Fourier transformation to the displacement as 
$\ul = \sum_{\qq}\uu_{\qq} e^{i\qq\cdot\Rl}$. 
The coefficient $\uu_{\qq}$ is written as 
$\uu_{\qq} = \sum_{\lambda} u_{\qq \lambda} \eqlam$ with the polarization vector $\eqlam$ for the mode $\lambda$. 
We label the longitudinal mode by $\lambda = {\rm l}$ and 
one of the transverse modes by $\lambda = {\rm t}$. 
Then, eq.~(\ref{del_v_c_exp}) is written as 
%(9)---------------------------------------------
\begin{align}
&\Delta \Vion = 
\frac{1}{2}\sum_{\qq,\GG,\lambda}u_{\qq \lambda}u_{-\qq \lambda}
A_{\lambda}(\GG, \qq) 
\label{del_v_c}
\end{align}
%---------------------------------------------
with $A_{\lambda}(\GG, \qq)$ = 
$\bar{v} (\GG+\qq) ((\GG+\qq)\cdot\eqlam)^2-$ 
$\bar{v} (\GG) (\GG\cdot\eqlam)^2$, 
where $\GG$ stands for the reciprocal lattice vector.  

Next, we calculate $\Delta\Vel$. 
The change in ionic configuration from $\{\Rl\}$ to $\{ \Rl + \ul \}$ changes the potential in eq.~(\ref{one_p_Hamiltonian}) by $\delta w$ $\equiv$ $w({\bf r}-\Rl-\ul)$$-$$w({\bf r}-\Rl)$. 
Accordingly, the transfer integral changes by 
$\delta t(\ulm)$ $=$ $t(\Rlm+\ulm)$$-$$t(\Rlm)$, 
which is on the same order as $\delta w$. 
The average of the electron operators 
in eq.~(\ref{electronic_pot}) changes as 
$\langle a_{\bfl\sigma}^\dagger a_{\bfm\sigma}\rangle _{\rm el}$ $=$ $\langle a_{\bfl\sigma}^\dagger a_{\bfm\sigma}\rangle _{\rm el}^{(0)}$ + $O((\delta w/\epsilon_{\rm F})^2)$, 
where $\langle\cdots \rangle_{\rm el }^{(0)}$ is the thermal average for $\{\Rl\}$ and $\epsilon_{\rm F}$ is the Fermi energy. 
Then, we obtain 
$\Delta\Vel=$ $(1/2)\sum_{\bfl,\bfm,\sigma} \delta t({\ulm})$
$(\langle a_{\bf l\sigma}^{\dagger}a_{\bf m\sigma}\rangle_{\rm el}^{(0)}+{\rm h.c.})$ 
in the first order of $\delta w/\epsilon_{\rm F}$ 
as an approximation. 
We manipulate $\Delta\Vel$ in the same manner as that for $\Vion$. 
Using the Fourier transform, $a_{\bfl\sigma} =$ $(1/\sqrt{N_L})$ 
$\sum_{\kk} a_{\kk\sigma} e^{i\kk\cdot\Rl}$, 
$t(\Rl)=(1/N_L)\sum_{\kk} t(\kk)e^{i\kk \cdot \Rl}$, and  $\bar{t}(\kk)=t(\kk)\exp[-\langle (\kk\cdot\uud)^2\rangle/2 ] $, we obtain
%(10)---------------------------------------------
\begin{align}
&\Delta \Vel 
= \frac{1}{2N_L}\sum_{\qq,\lambda}u_{\qq\lambda}u_{-\qq\lambda} \sum_{\kk,\sigma} f_{\kk}
B_{\lambda}(\kk, \qq) 
\label{del_v_e}
\end{align}
%---------------------------------------------
with $B_{\lambda}(\kk, \qq)$ = 
$\bar{t} (\kk+\qq) ((\kk+\qq)\cdot\eqlam)^2-$
$\bar{t} (\kk) (\kk\cdot\eqlam)^2$. 

Using eqs. (\ref{del_v_c}) and (\ref{del_v_e}), the Hamiltonian (\ref{Hamiltonian}) exhibits a bilinear form. 
Hence, the energy $\omega_{\qq \lambda}$ of the normal vibration exhibits the dispersion 
%(11)---------------------------------------------
\begin{align}
M\omega_{\qq \lambda}^2 &= \sum_{\GG} 
A_{\lambda}(\GG, \qq) 
+\frac{2}{N_L}\sum_{\kk} f_{\kk} 
B_{\lambda}(\kk, \qq) . 
\end{align}
%---------------------------------------------
The longitudinal velocity $c_\ql$ and transverse velocity $c_\qt$ are determined using 
$\omega_\ql=c_\ql |\qq|$ ($\eql\parallel \qq$) and 
$\omega_\qt=c_\qt |\qq|$ ($\eqt\perp \qq$) 
for a small momentum $\qq$, respectively. 
Then, the longitudinal velocity is given as 
%(12)---------------------------------------------
\begin{align}
M c_\ql^2 = v(0) 
+ \sum_{\GG\neq 0} \Gamma_\ql(\GG) \bar{v} (\GG) 
+\frac{2}{N_L}\sum_{\kk}f_{\kk} \Gamma_\ql(\kk) \bar{t} (\kk) 
\label{velocity_q_l} 
\end{align}
%---------------------------------------------
with the operator $\Gamma_\ql(\KK) = 1
+2(\eql\cdot\KK)(\eql\cdot\nabla_{\KK})
+(1/2)(\eql\cdot\KK)^2(\eql\cdot\nabla_{\KK})^2$
for $\KK = \GG$ or $\kk$. 
The transverse velocity is given as 
%(13)---------------------------------------------
\begin{align}
M c_\qt^2 = \sum_{\GG\neq 0}
\Gamma_\qt(\GG) \bar{v} (\GG)
+\frac{2}{N_L}\sum_{\kk}f_{\kk}
\Gamma_\qt(\kk) \bar{t}(\kk) 
\label{velocity_q_t}
\end{align}
%---------------------------------------------
with the operator $\Gamma_\qt(\KK) = (1/2)(\eqt\cdot\KK)^2(\eql\cdot\nabla_{\KK})^2$ for $\KK = \GG$ or $\kk$. 
In the limit of $\qq \rightarrow 0$, we simply denote the velocities as $\cl$ and $\ct$. 
The velocities depend on the renormalized ionic potential $\bar{v}$ and the transfer integral $\bar{t}$, which include $\cl$ and $\ct$ in reduction factors. 
Hence, we obtained a set of self-consistent equations (\ref{velocity_q_l}) and (\ref{velocity_q_t}) to determine $\cl$ and $\ct$ as functions of temperature. 
As temperature increases, $\bar{v}$ and $\bar{t}$ decrease and then $\cl$ and $\ct$ decrease. 
We find that $\ct$ vanishes at a special temperature, which we identify the melting temperature. 
On the other hand, the longitudinal velocity $\cl$ remains finite because eq.~(\ref{velocity_q_l}) includes the ionic potential $v(0)$, which is independent of temperature. 
Actually, the longitudinal phonon remains in the liquid phase. 

We show that near the melting point, 
the ionic potential $\bar{v}(\GG)$ for $\GG\neq 0$ becomes 
much smaller than the transfer integral $\bar{t}(\kk)$ 
for sc, bcc and fcc crystals. 
The Fourier transforms of the ionic potential $\bar{v} (\GG)$ and 
the transfer energy $\bar{t}(\kk)$ are given as
%(14)(15)---------------------------------------------
\begin{align} 
 \bar{v} (\GG)&=v(\GG)\langle e^{-i\GG\cdot\uud}\rangle =v(\GG)e^{-\frac{1}{6}\GG^2\langle \uud^2\rangle }, 
\label{v_bar}
\\
\bar{t}(\kk)&=t(\kk)\langle e^{-i\kk\cdot\uud}\rangle 
=t(\kk)e^{-\frac{1}{6}\kk^2\langle \uud^2\rangle }. 
\label{t_bar}
\end{align}
%---------------------------------------------
Here, the mean square amplitude of the vibration between the neighboring sites is calculated as 
%(16)---------------------------------------------
\begin{align}
\langle \uud^2\rangle =\sum_{\qq}\left(2\sin\frac{\qq\cdot\mib{R}_{\rm d}}{2}\right)^2\langle \uu_{\qq} \uu_{-\qq}\rangle , 
\label{msa_vib}
\end{align}
%---------------------------------------------
where $\mib{R}_{\rm d}$ is the nearest-neighbor lattice coordinate and the last factor is calculated as 
%(*)---------------------------------------------
\begin{align*} 
\langle \uu_{\qq} \uu_{-\qq}\rangle 
=\frac{1}{N_LM} \left\{ \frac{2}{\ct q} \ 
g\!\left(\frac{\ct q}{\kB T}\right)
+\frac{1}{\cl q} \ 
g\!\left(\frac{\cl q}{\kB T}\right)
\right\} 
\end{align*}
%---------------------------------------------
with the function $g(x) = (e^x - 1)^{-1} + 1/2$. 
At high temperatures, eq.~(\ref{msa_vib}) reduces to 
%(17)---------------------------------------------
\begin{align}
 \langle \uud^2\rangle = 
 \frac{\kB T\Rd^2}{3M}\left( \frac{2}{\ct^2}+\frac{1}{\cl^2} \right) 
\end{align}
%---------------------------------------------
on the assumption of $\cl q/\kB T$, $\ct q/\kB T\ll 1$. 
We compare two reduction factors in eqs. (\ref{v_bar}) and (\ref{t_bar}). 
We take the magnitude $G_{\rm min}$ of one of the smallest reciprocal lattice vectors for the lower bound of $|\GG|$ in eq.~(\ref{v_bar}) 
and the Fermi wave number $\kF$ for the upper bound of $|\kk|$ in eq.~(\ref{t_bar}). 
Then, the ratio $r(\nee) \equiv G^2_{\rm min}/\kF^2$ measures the relative importance of $\bar{v}(\GG)$ against $\bar{t}(\kk)$, where $\nee$ is the electron number per site. 
For a simple cubic lattice, the ratio is written as 
$r(\nee) = ({8\pi }/{3\nee}) ^{2/3}$, 
which gives $r(1) = 4.12$ and $r(2) = 2.60$. 
For a bcc lattice, the ratio is written as 
$r(\nee) = 2({4\pi }/{3\nee}) ^{2/3}$, 
which gives $r(1) = 5.20$ and $r(2) = 3.28$. 
For an fcc lattice, the ratio is written as 
$r(\nee) = 3({2\pi }/{3\nee}) ^{2/3}$, 
which gives $r(1) = 4.91$ and $r(2) = 3.09$. 
These estimations show that $\bar{v}(\GG)$ becomes negligible in comparison with $\bar{t}(\kk)$. 
Therefore, near the melting point, we reasonably drop the first term in eq.~(\ref{velocity_q_t}) and obtain 
%(18)---------------------------------------------
\begin{align}
c_\qt^2=
\frac{1}{MN_L}\sum_{\kk}f_{\kk} 
\left\{ (\kk\cdot\eqt)^2
\left( \eql\cdot\nabla_{\kk} \right)^2
\, \bar{t}(\kk) \right\} . 
\label{c_integral}
\end{align}
%---------------------------------------------
This equation implies that the melting temperature is determined solely by the force constant given by the attractive potential due to conduction electrons. 

%-->-->-->-->-->-->-->-->-->-->-->-->-->-->-->-->
%Fig. 1
\begin{figure}[t] 
\begin{center}\leavevmode
\includegraphics[width=0.85\linewidth]{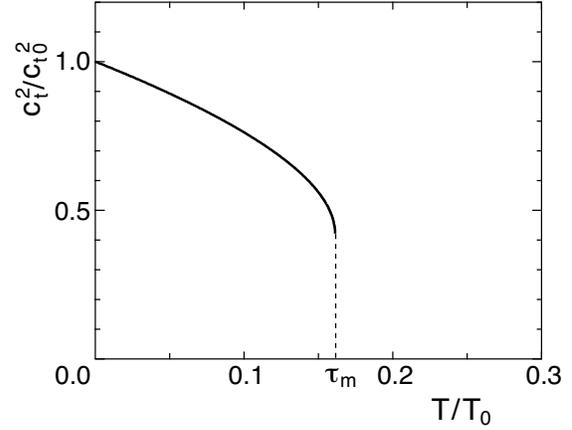}
\end{center}
\caption{${\ct^2}/{\ctzero^2}$ versus $T/T_0$. 
$\tau_{\rm m}=0.161121$ is the melting temperature scaled by $T_0$.} 
\label{velocity}
\end{figure}
%--<--<--<--<--<--<--<--<--<--<--<--<--<--<--<--<

Since we have used the nearly free electron model, we have 
$t(\kk) = \hbar^2\kk^2/(2m^*)$ 
with the effective electron mass $m^*$. 
When the sound propagates along the $x$-direction with the polarization vector parallel to the $y$-direction, eq.~(\ref{c_integral}) becomes
%(19)---------------------------------------------
\begin{align}
\ct^2=\frac{V}{(2\pi)^3MN_L} 
\int f_{\kk} \, k_y^2 \, 
\frac{\partial^2}{\partial k_x^2}
\left(\frac{\hbar^2 k^2}{2m^*}e^{-\kappa k^2}\right)
d^3\kk 
\label{c_t_integral}
\end{align}
%---------------------------------------------
with $\kappa=\Rd^2\kB T/(9M\ct^2)$ under the assumption of $\ct \ll \cl$. 
By defining the constants 
$T_0 = 9\hbar^2 \nee / (10m^*\Rd^2\kB)$ and 
$\ctzero = \hbar \kF \nee^{1/2} / (10Mm^*)^{1/2}$ 
and introducing the scaled variables 
$\gamma={\ct^2}/{\ctzero^2}$ and $\tau=T/T_0$, 
eq.~(\ref{c_t_integral}) reduces to 
%(20)---------------------------------------------
\begin{align}
\gamma= \left( 1-\frac{\tau}{\gamma} \, \right) 
e^{-\tau/\gamma} 
\label{gamma_tau}
\end{align}
%---------------------------------------------
after integration. 
Note that this relation is universal, and in particular, 
$M$ is absorbed in the scaled velocity $\sqrt{\gamma}$. 

By introducing a variable $x \equiv \tau/\gamma$, eq.~(\ref{gamma_tau}) is written as 
$\tau = x(1-x)e^{-x}$. 
The maximum $\tau$, $\tau_{\rm m}$, is determined by $d\tau/dx = 0$. 
We denote $x$ and $\gamma$ for $\tau = \tau_{\rm m}$ as $x_{\rm m}$ and $\gamma_{\rm m}$, respectively. 
These values are given as 
%(21)---------------------------------------------
\begin{equation}
\begin{split}
x_{\rm m} &= (3-\sqrt{5})/2=0.381966, \\
\tau_{\rm m} &= x_{\rm m}(1-x_{\rm m})e^{-x_{\rm m}}=0.161121 , \\
\gamma_{\rm m} &= \tau_{\rm m} / x_{\rm m} =0.421819. 
\end{split}
\end{equation}
%--------------------------------------------- 
We show $\gamma$ as a function of $\tau$ in Fig.~\ref{velocity}. 
We see that, as $T$ increases, $\ct$ decreases and jumps down to zero at $T_{\rm m}=\tau_{\rm m}T_0$. 
Thus, we arrive at eq.~(\ref{melting_temp}), where 
$T_{\rm m} = 0.145009 \, \hbar^2 \nee/(m^*\Rd^2\kB).$ 

%-->-->-->-->-->-->-->-->-->-->-->-->-->-->-->-->
%Table 1
\begin{table}[t]
\begin{center}
\begin{tabular}{ccccccccc}\hline
&$\nee$&$\displaystyle\frac{m^{*}}{m}$&$\Rd ({\rm \AA})$
&$T_{\rm m}^{\rm (th)} ({\rm K})$&$T_{\rm m}^{\rm (ex)} ({\rm K})$&
$\displaystyle\frac{T_{\rm m}^{\rm (th)}}{T_{\rm m}^{\rm (ex)}}$
\\ \hline
Li (bcc) & 1 & 2.18 & 3.023  & 643.7 & 453.7 & 1.43\\ 
Na (bcc) & 1 & 1.26 & 3.659  & 760.1 & 371.0 & 2.05\\ 
K  (bcc) & 1 & 1.25 & 4.525  & 501.0 & 336.3 & 1.49\\ 
Rb (bcc) & 1 & 1.26 & 4.837  & 435.0 & 312.6 & 1.39\\ 
Cs (bcc) & 1 & 1.43 & 5.235  & 326.9 & 301.6 & 1.08\\ 
\hline
Cu (fcc) & 1 & 1.38 & 2.56   & 1418  & 1358 & 1.04\\ 
Ag (fcc) & 1 & 1.00 & 2.89   & 1535  & 1235 & 1.24\\ 
Au (fcc) & 1 & 1.14 & 2.88   & 1356  & 1338 & 1.01\\ 
\hline
Ca (fcc) & 2 & 1.90 & 3.95   & 865.1 & 1113 & 0.777\\ 
Sr (fcc) & 2 & 2.00 & 4.30   & 693.5 & 1042 & 0.666\\ 
Ba (bcc) & 2 & 1.40 & 4.35   & 968.1 & 1002 & 0.966\\ 
\hline
Al (fcc) & 1 & 1.48 & 2.86 &1058.2 & 933.5 & 1.13\\ 
\hline
\end{tabular}
\end{center}
\caption{Comparison of the theoretical melting temperature $T_{\rm m}^{\rm (th)}$ by eq.~(\ref{melting_temp}) to the experimental one $T_{\rm m}^{\rm (ex)}$ for various metals. 
Data for $\nee$, $m^{*}$, $\Rd$, and $T_{\rm m}^{\rm (ex)}$ are referred from 
ref.~\citen{Kittel}.}
\label{Tm}
\end{table}
%--<--<--<--<--<--<--<--<--<--<--<--<--<--<--<--<

In Table~\ref{Tm}, we compare $T_{\rm m}^{({\rm th})}$, which is the melting temperature value calculated from eq.~(\ref{melting_temp}), to $T_{\rm m}^{({\rm ex})}$, which is that from experiments, for various metals. 
The metals have monovalent and divalent elements with conductive s-electrons and with cubic symmetry. 
For $m^*$, $\nee$ and $\Rd$, we rely on Kittel's textbook~\cite{Kittel}. 
For Al, we {replace} $\nee=3$ with $\nee=1$, since the Hall coefficient shows that the carrier in Al is a single hole~\cite{Kittel}.
Note that variations in $M$ are typically represented by $M({\rm Cs})/M({\rm Li}) \simeq$ 19.2 in alkali metals and by $M({\rm Au})/M({\rm Cu}) \simeq$ 3.1 in noble metals. 
In contrast, the corresponding variations in $T_{\rm m}^{\rm (ex)}$ in Table~\ref{Tm} are relatively small. 
Hence, the approximations used to derive the $M$-independent equation (\ref{melting_temp}) are considered to be reasonable. 

We see that $T_{\rm m}^{\rm (th)}$ agrees with $T_{\rm m}^{\rm (ex)}$ in accuracy within 10~\% (Cs, Cu, Au, and Ba) and within 50~\% (all metals in Table~\ref{Tm} except Na).  
In noble metals, the agreement is good, but in alkali metals, $T_{\rm m}^{\rm (th)}$ is somewhat higher than $T_{\rm m}^{\rm (ex)}$.
The general equation (\ref{melting_temp}) gives definite values for the melting temperatures of metals without adjustable parameters. 
Although we find a certain numerical difference between $T_{\rm m}^{\rm (th)}$ and $T_{\rm m}^{\rm (ex)}$, 
the values in Table~\ref{Tm} are considered to be theoretical bases for further development of the study of melting temperature. 

Now, we evaluate the Lindemann ratio in our theory. 
The mean square amplitude of vibration at high temperatures is
%(22)--------------------------------------------- 
\begin{align}
\langle {u}_0^2\rangle =\frac{1}{3}\sum_{\qq} \langle \uu_{\qq}\uu_{-\qq}\rangle =\alpha\frac{\kB T_{\rm m}}{M}\left( \frac{2}{\ct^2}+\frac{1}{\cl^2} \right) 
\label{vibration_high}
\end{align}
%--------------------------------------------- 
with $\alpha=(1/3N_L)\sum_{\qq} (1/q^2)$. 
By neglecting $1/\cl^2$, we obtain the Lindemann ratio $\delta$ shown as 
%(23)--------------------------------------------- 
\begin{align}
\delta^2 \equiv 
\frac{\langle {u}_0^2\rangle }{\Rd^2}\sim \frac{\alpha}{\Rd^2}\frac{\kB T_{\rm m}}{M}\frac{2}{\ct^2}
=\frac{\alpha}{\Rd^2}\frac{18 x_{\rm m}}{(\kF\Rd)^2} . 
\end{align}
%--------------------------------------------- 
With 
$\alpha \Rd^{-2}$ = 0.05529 (bcc) and 0.05224 (fcc), and with $18/(\kF\Rd)^2$ = 1.580 (bcc) and 1.493 (fcc), we obtain 
%(24)--------------------------------------------- 
\begin{align}
\delta = 0.183 \ ({\rm bcc}) \quad {\rm and} \quad 0.172 \ ({\rm fcc}). 
\end{align}
%--------------------------------------------- 
These values are consistent with and then  confirm the Lindemann criterion $\delta \sim 0.1$. 
Our results also agree with $\delta\simeq 0.15 $ and 0.126 obtained by numerical calculations with the soft core repulsive potential $r^{-n}$ $(n>3)$ and the hard core repulsive potential, respectively~\cite{Alder3,Hoover3}. 

In summary, we use the shear instability of the solid phase to derive the equation for the melting temperature. 
We renormalize the ionic and electronic restoring forces, including higher order terms of displacement enhanced near the melting temperature by applying SCHA. 
The ionic force due to the renormalized ionic potential  $\bar{v}(\GG)$ with $\GG\neq 0$ decreases significantly and becomes negligible near the melting temperature. Hence, the velocity $\ct$ is solely determined by the force via conduction electrons. 
We obtain the melting temperature from the vanishing point of velocity $\ct$ 
without any adjustable parameters. 
The theoretical melting temperature $T_{\rm m}^{\rm (th)}$ is consistent with the experimental one $T_{\rm m}^{\rm (ex)}$. 
We can estimate the melting temperatures of various pure and composite metals if we have the values of  $m^*, \Rd$ and $\nee$. 
The Lindemann criterion is obeyed. 
Finally, we discuss melting temperatures of the transition, lanthanide, and actinide metals. 
Using the experimental data of $\Rd$, we obtain $T_{\rm m}^{\rm (th)}/T_{\rm m}^{\rm (ex)}=A(\nee/m^*)$: 
$A$=0.7$\sim$1.2 (3{\rm d}), 0.6$\sim$0.9 (4{\rm d}), and 0.8$\sim$0.5 (5{\rm d}) for the transition metals; 
$A$=0.7$\sim$0.9 for the lanthanide metals; 
and  $A$=0.5$\sim$0.7 (Th, Pa, and Am) for the actinide metals. 
If $\nee/m^*$ is about $1 \sim 2$, 
$T_{\rm m}^{\rm (th)}/T_{\rm m}^{\rm (ex)}\sim 1$. 
Actually, $\nee/m^*\sim 2$ in the lanthanide and actinide metals, and the transition metals may share similar values. 

\section*{Acknowledgement}
We would like to thank I. Kawabe and B. Martyr for discussion.

\end{document}